\documentclass[sigconf,nonacm]{acmart}
\AtBeginDocument{%
  \providecommand\BibTeX{{%
    \normalfont B\kern-0.5em{\scshape i\kern-0.25em b}\kern-0.8em\TeX}}}

\usepackage{savesym}
\savesymbol{zifour@default}
\savesymbol{zifour@scaled}
\usepackage{enumerate}
\usepackage{multirow} 
\usepackage{array}
\usepackage{cleveref}
\usepackage{inconsolata}
\usepackage{spverbatim}
\usepackage{wrapfig}

\usepackage{graphicx}
\usepackage{makecell}
\usepackage{xspace}

\newcommand{\comm}[1]{}

\newcommand{\revdel}[1]{}

\newcommand{\denselist}{\itemsep 0pt\parsep=0pt\partopsep 0pt\vspace{-\topsep}}

\newcommand{\declarativebaseline}{DeclBase}

\begin{document}

\title
[Imperative vs. Declarative Programming Paradigms for Open-Universe Scene Generation]
{Imperative vs. Declarative Programming Paradigms\\for Open-Universe Scene Generation}

\author{Maxim Gumin}
\email{maxim_gumin@brown.edu}
\affiliation{%
    \institution{Brown University}
    \country{USA}
}

\author{Do Heon Han}
\email{do_heon_han@brown.edu}
\affiliation{%
    \institution{Brown University}
    \country{USA}
}

\author{Seung Jean Yoo}
\email{seung_jean_yoo@brown.edu}
\affiliation{%
    \institution{Brown University}
    \country{USA}
}

\author{Aditya Ganeshan}
\email{aditya_ganeshan@brown.edu}
\affiliation{%
    \institution{Brown University}
    \country{USA}
}

\author{R. Kenny Jones}
\email{russell_jones@brown.edu}
\affiliation{%
    \institution{Brown University}
    \country{USA}
}

\author{Rio Aguina-Kang}
\email{raguinakang@ucsd.edu}
\affiliation{%
    \institution{UC San Diego}
    \country{USA}
}

\author{Stewart Morris}
\email{stewart_morris@brown.edu}
\affiliation{%
    \institution{Brown University}
    \country{USA}
}

\author{Daniel Ritchie}
\email{daniel\_ritchie@brown.edu}
\affiliation{%
    \institution{Brown University}
    \country{USA}
}

\begin{abstract}
Current methods for generating 3D scene layouts from text predominantly follow a declarative paradigm, where a Large Language Model (LLM) specifies high-level constraints that are then resolved by a separate solver. This paper challenges that consensus by introducing a more direct, imperative approach. We task an LLM with generating a step-by-step program that iteratively places each object relative to those already in the scene. This paradigm simplifies the underlying scene specification language, enabling the creation of more complex, varied, and highly structured layouts that are difficult to express declaratively. To improve the robustness, we complement our method with a novel, LLM-free error correction mechanism that operates directly on the generated code, iteratively adjusting parameters within the program to resolve collisions and other inconsistencies. In forced-choice perceptual studies, human participants overwhelmingly preferred our imperative layouts, choosing them over those from two state-of-the-art declarative systems 82\% and 94\% of the time, demonstrating the significant potential of this alternative paradigm. Finally, we present a simple automated evaluation metric for 3D scene
layout generation that correlates strongly with human judgment.

\end{abstract}



\begin{teaserfigure}
    \centering
    \setlength{\tabcolsep}{2pt}
    \newcolumntype{y}{>{\centering\arraybackslash}p{0.24\linewidth}}
    \begin{tabular}{yyyy}
         \textit{``A casino''} &
         \textit{``Forest clearing''} &
         \textit{``Depths of hell''} &
         \textit{``A post-apocalyptic campsite''}
         \\
         \includegraphics[width=\linewidth]{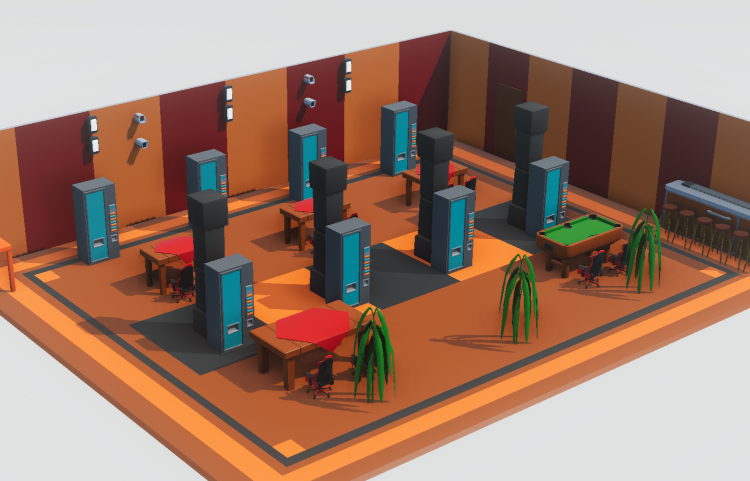} &
         \includegraphics[width=\linewidth]{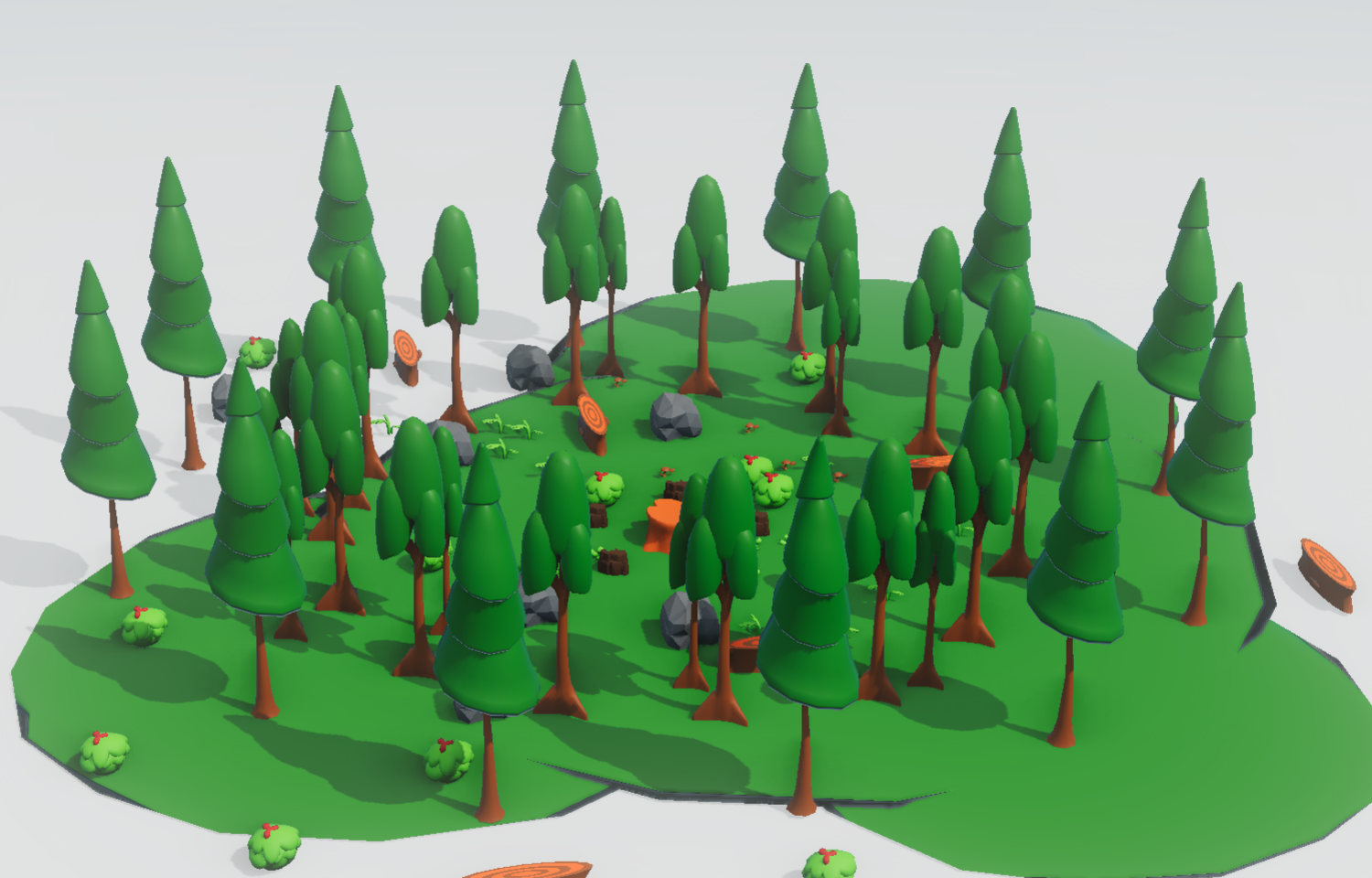} &
         \includegraphics[width=\linewidth]{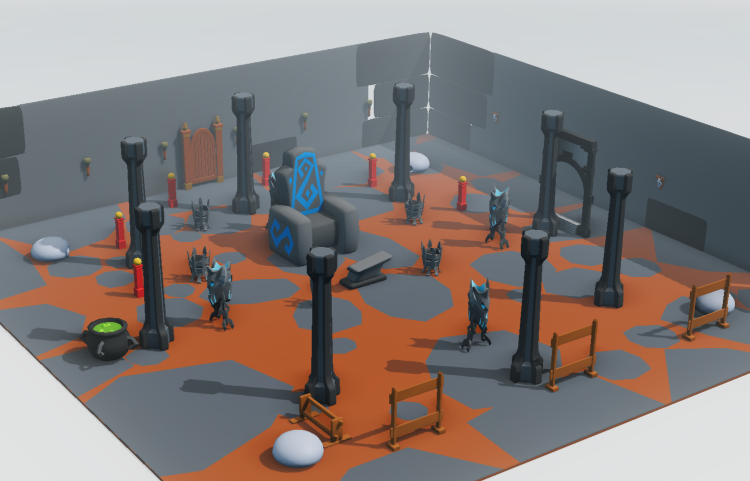} &
         \includegraphics[width=\linewidth]{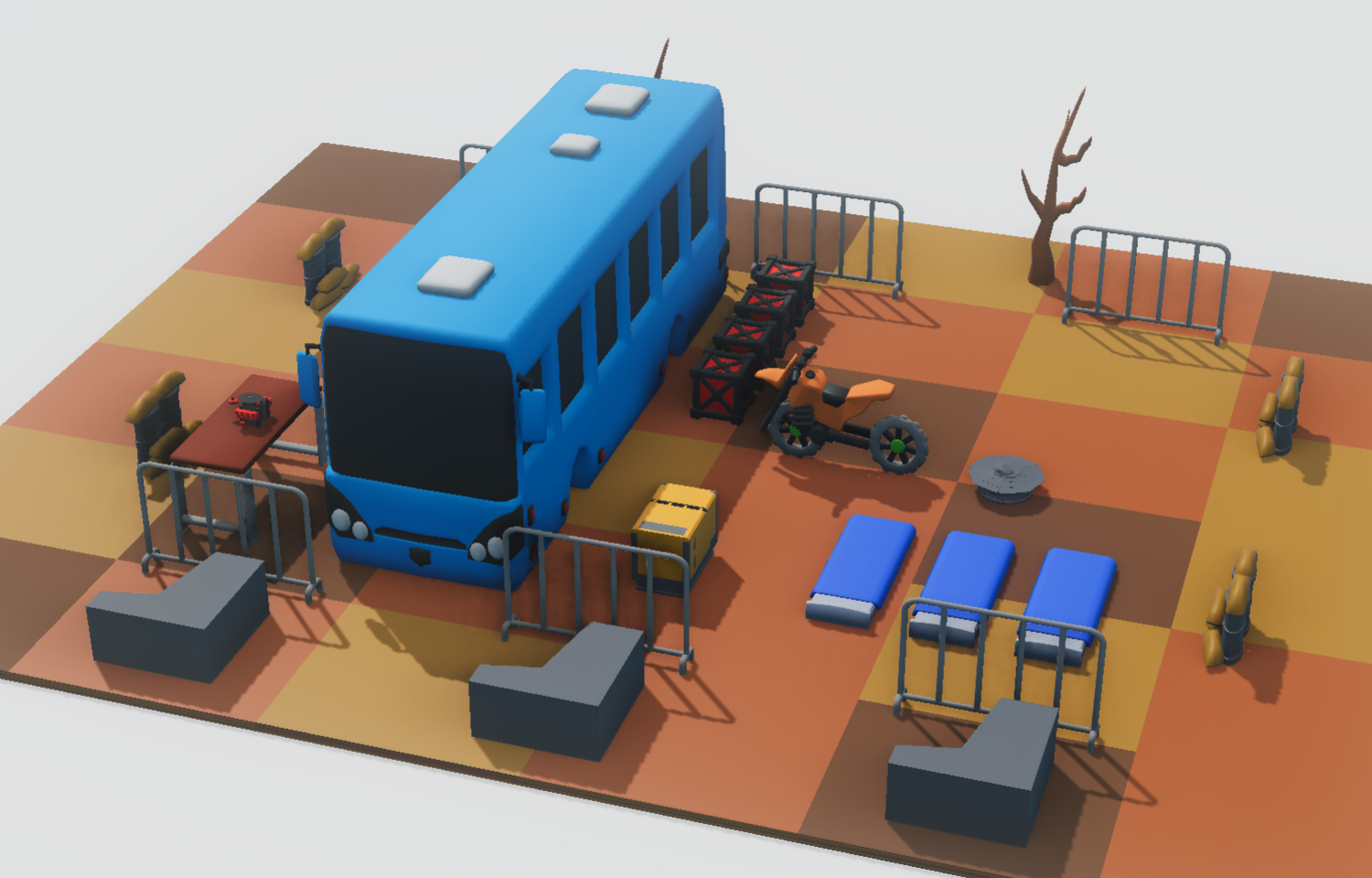}
         \\
         \textit{``Stonehenge''} &
         \textit{``An ice city''} &
         \textit{``A classroom''} &
         \textit{``A graveyard''}
         \\
         \includegraphics[width=\linewidth]{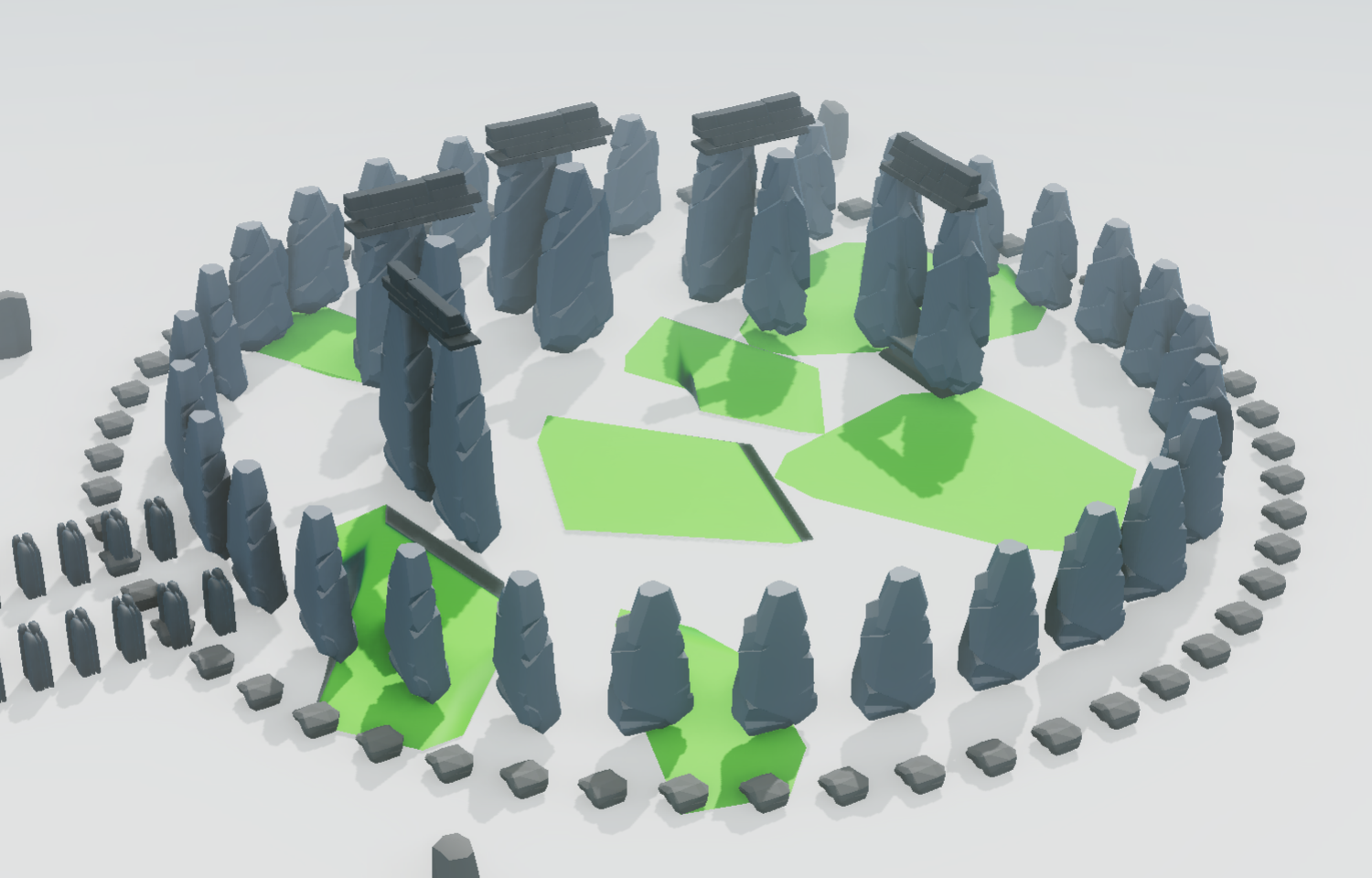} &
         \includegraphics[width=\linewidth]{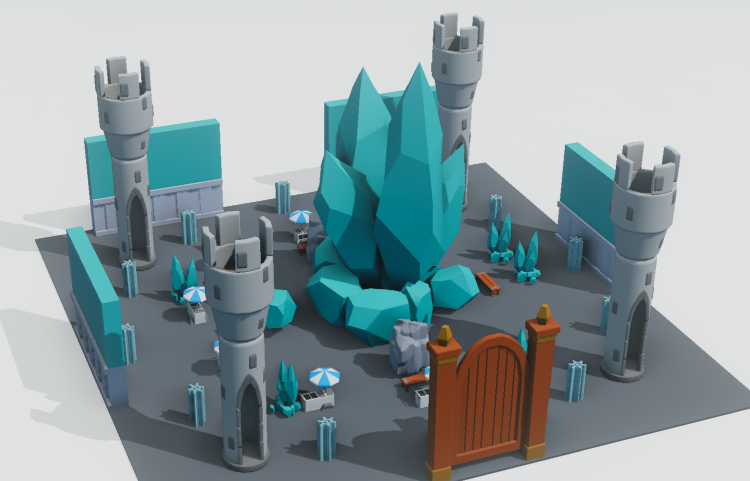} &
         \includegraphics[width=\linewidth]{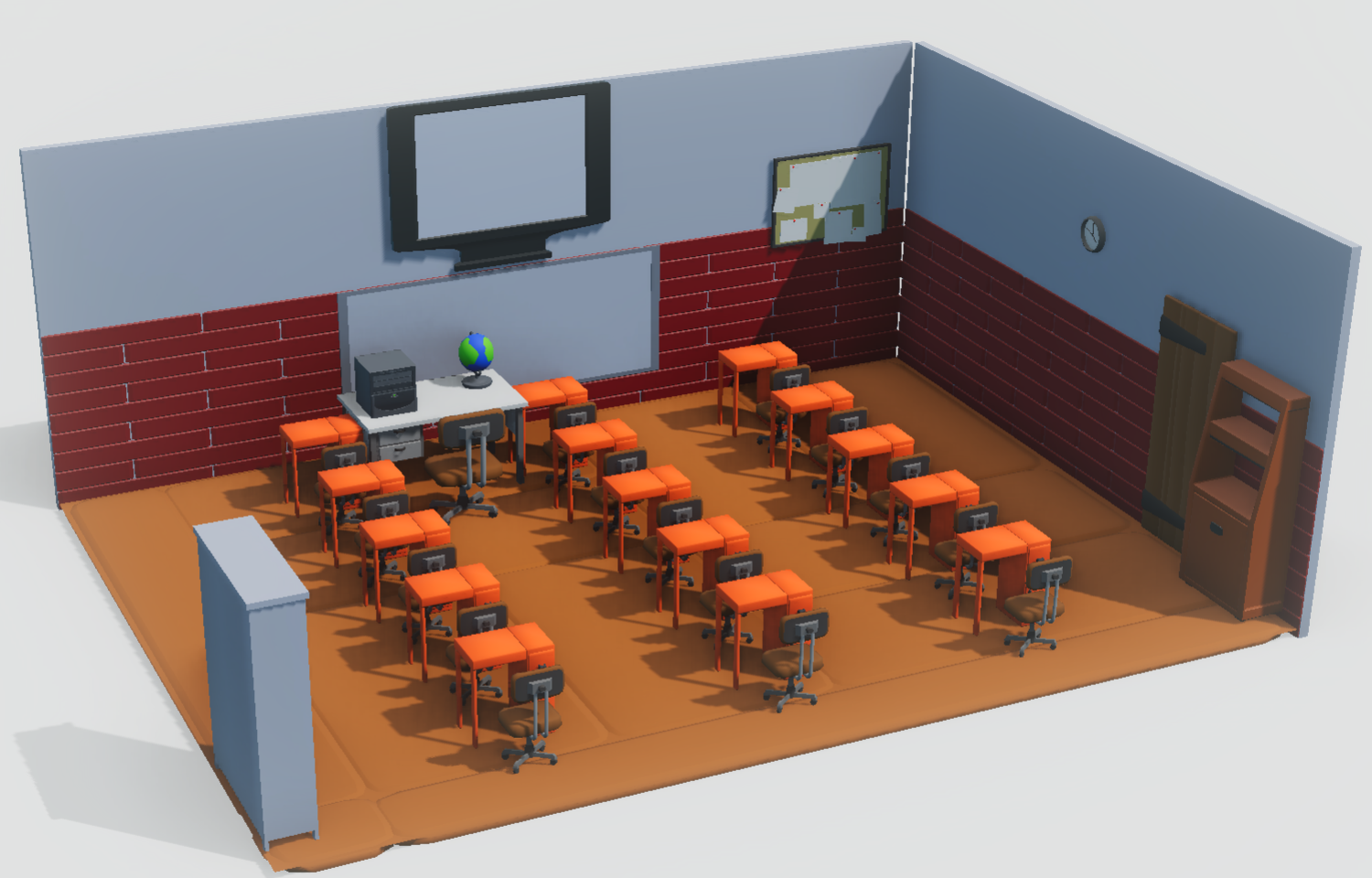} &
         \includegraphics[width=\linewidth]{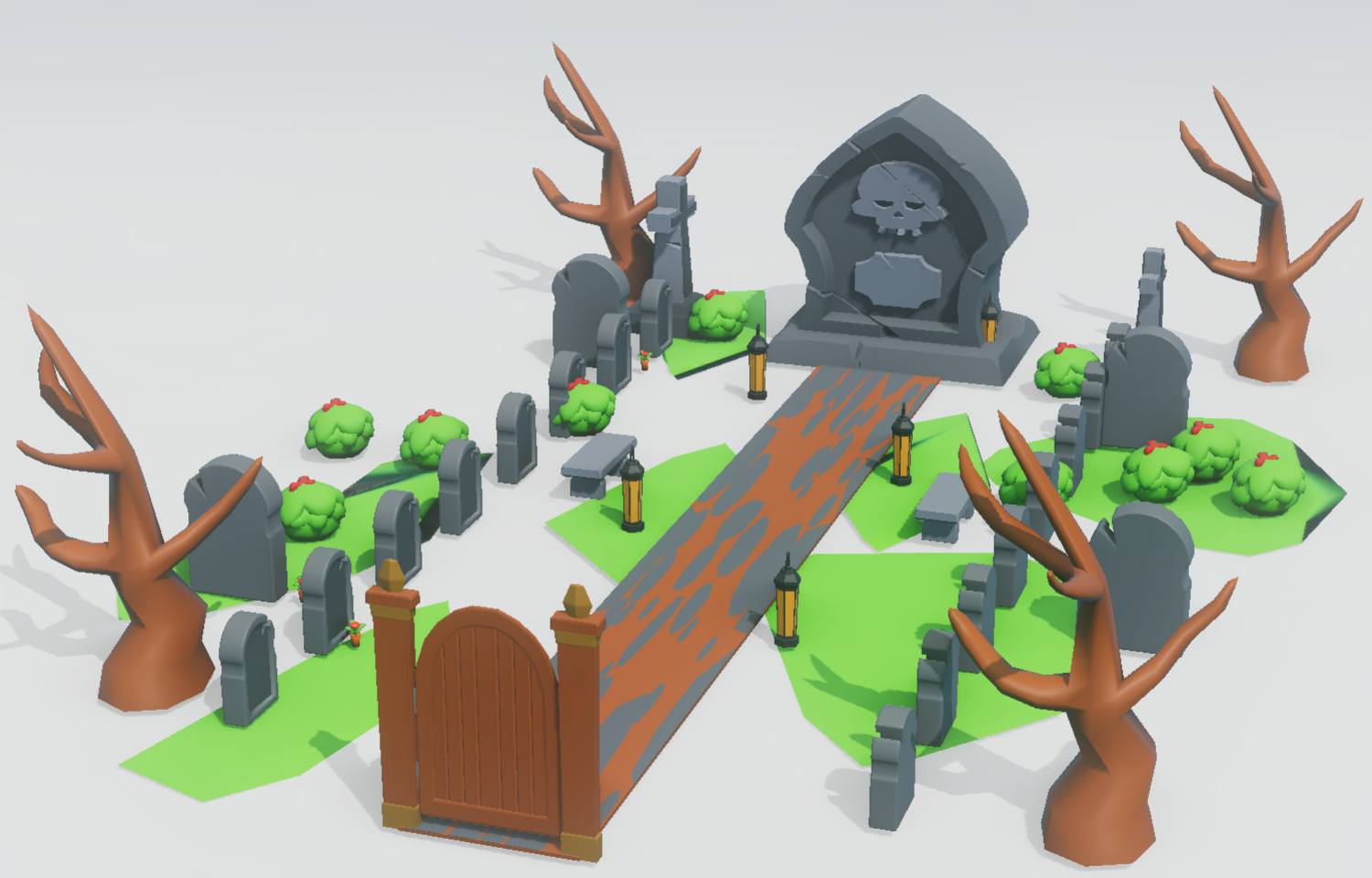}
    \end{tabular}
    \caption{
        Our method generates 3D indoor and outdoor scenes from open-ended text prompts.        Generated scenes are not limited to a fixed set of room types or object categories.
        All scenes in the figure are generated using our imperative approach with error correction mechanism.
    }
    \label{fig:teaser}
\end{teaserfigure}

\maketitle

\section{Introduction}
3D scenes serve as representations of the environments surrounding us: homes, workplaces, social gathering spaces, etc. They can also represent virtual worlds for games, films, and architecture.
In this paper, we address \emph{open-universe scene generation}: synthesizing a 3D scene from a natural language prompt, where prompts are not limited to a fixed vocabulary, and objects are not restricted to a fixed set of object categories. Large language models (LLMs) are a natural fit for this task given their vast knowledge bases.

\begin{figure*}[t!]
    \centering
    \includegraphics[width=0.8\linewidth]{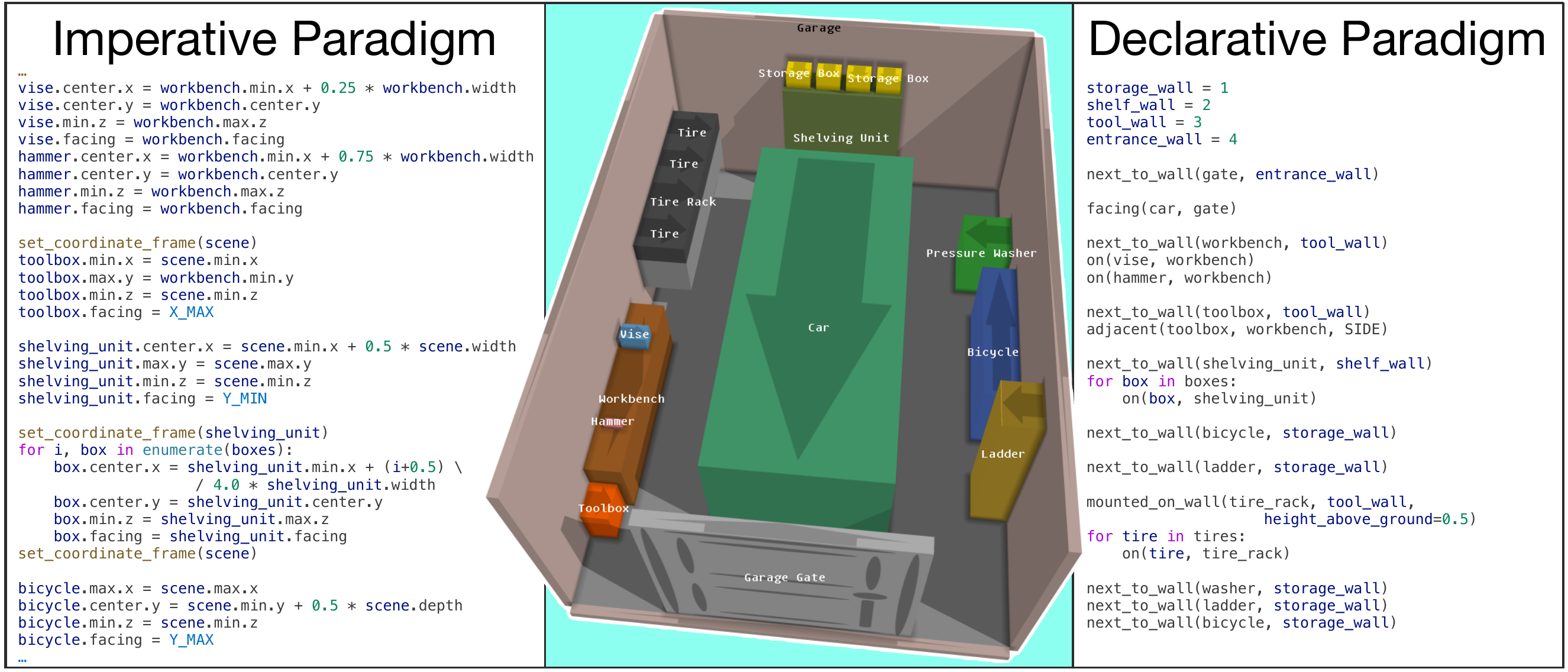}
    \caption{
    Comparison of the commonly used declarative paradigm (right) and our proposed imperative paradigm (left) for generating a "Garage" scene layout. The imperative paradigm explicitly specifies geometric relationships between objects, enabling flexible and precise arrangements. 
    }
    \label{fig:code-compare}
\end{figure*}

A 3D scene can be viewed as a collection of objects, where each object is specified by attributes such as size, mesh, position, and orientation.
Synthesizing such a scene involves several steps: generating a set of objects, determining the positions and orientations of those objects (i.e. layout), and generating or retrieving 3D meshes for each object.
In this paper, we focus on the layout subproblem.


For the task of layout generation, recent work has overwhelmingly adopted what we call the \emph{declarative} paradigm. In this paradigm, the LLM does not synthesize explicit object coordinates. Instead, it synthesizes a set of relations between objects, such as \texttt{on(a,b)}, \texttt{adjacent(a,b)} or \texttt{aligned(a,b,c)}. Then, a solver module finds a configuration of object positions that satisfies all the relations (or as many relations as possible). 
The rationale behind the declarative paradigm is that it should be easier for an LLM to reason about sentences such as “the lamp is on the table” or “the chair is adjacent to the table” than about precise numeric values.



While the declarative paradigm can work for describing small household scenes such as bedrooms and living rooms, it can struggle with highly structured scenes (e.g. courtrooms, grocery stores) and large outdoor scenes (e.g. forests, city blocks). 
Our intuition is that such scenes require geometric relations that are not expressible in the given declarative domain specific language (DSL). More relations can be added to the DSL, but adding new relations comes at a price, as the DSL becomes more complex, necessitating long, complicated input prompts and in-context examples for the LLM, which may result in errors.
Declarative approaches may also struggle with scenes containing a large number of objects (e.g. museums, theaters), because the time required for a solver to find a satisfying configuration of object positions depends at least linearly on the number of relations, and the number of relations usually depends quadratically on the number of objects.

An alternative to the declarative paradigm is the \emph{imperative} paradigm, in which objects are placed one at a time, with each new object positioned relative to those already in the scene.
We illustrate the scene programs for a ``Garage'' scene under this paradigm in Figure~\ref{fig:code-compare} (left). 
As shown in the figure, this paradigm allows us to (a) explicitly position objects in reference to others (for instance, the \texttt{vise} and \texttt{workbench} are positioned in relation to the \texttt{workbench}), and (b) specify nuanced parameterized geometric relations between objects (as done for arranging \texttt{boxes} on the \texttt{shelving\_unit}).


However, LLMs can still make errors in imperative scene programs. For example, an LLM might place a chair to the left of the table with insufficient space between the table and the wall to fit a chair. While a constraint solver in the declarative paradigm could solve for appropriate positions of both the chair and table, the imperative system cannot reposition the table once it is placed.

To address this challenge, we introduce an \emph{error correction procedure} that iteratively refines LLM-generated programs while preserving their original structure as much as possible. Using a coordinate descent-inspired approach, the mechanism adjusts one scene parameter at a time to reduce layout errors, ensuring minimal deviation from the initial program. This process is particularly effective for imperative scenes due to the presence of shared variables, which allow coordinated updates across multiple objects. For example, modifying a shared spacing parameter can automatically adjust the layout of an entire row of objects, maintaining structural consistency and avoiding overlaps. By refining programs in this manner, the correction procedure enhances the robustness of imperative scene generation without requiring additional LLM calls.

We compare our imperative approach with error correction against declarative scene layout generation methods on a variety of different scene prompts, ranging from small indoor scenes (e.g. Living Room, Classroom) to large outdoor scenes (e.g. Forest Clearing, City Block), from common spaces (e.g. College Gym, Parking Lot) to fantastical scenes (e.g. Depths of Hell, Skyward Kingdom), and from chaotic (e.g. Kindergarten, Post-Apocalyptic Campsite) to highly structured (e.g. Courtroom, Railway Station Platform) scenes. Contrary to the prevailing belief in the field, the imperative paradigm, if equipped with the error correction procedure, is competitive with the declarative paradigm. In forced-choice perceptual studies, human participants prefer scenes generated with our imperative system 82\% of times against a strong declarative baseline and 94\% of times against the declarative Holodeck~\cite{yang2023holodeck} system.

As perceptual studies can be time-consuming and potentially costly, we also investigated automated metrics for evaluating generated scene layouts.
We found that metrics designed to evaluate systems for other types of text-based visual generation~\cite{lin2024evaluating,JaeminCho2024} aligned poorly with human preferences from our perceptual study.
Thus, we introduce a new automated method for evaluating scene layout generation systems.
Our method takes a text prompt and a pair of scenes as input and outputs which of the two scenes is a better realization of the text prompt. 
Our method is straightforward to implement, using only a single call to a multimodal LLM.
Our approach achieves over 77\% agreement with human preferences, whereas existing automated metrics perform only marginally better than chance.


In summary, our contributions are:
\begin{enumerate}[(i)]
    \denselist
    \item Developing the imperative paradigm for open-universe scene generation and comparing it against the best systems that follow the declarative paradigm.
    \item An error correction method for imperative scene programs that does not involve additional calls to an LLM.
    \item A protocol for evaluating open-universe scene layout synthesis systems, including a benchmark set of input descriptions covering a wide variety of possible scenes.
    \item A human-aligned automated evaluation method for scene layout generation.
\end{enumerate}

Our code will be made available as open source upon publication.

\section{Related Work}
\paragraph*{Scene Synthesis pre-LLMs}
The problem of scene synthesis has a rich history in computer graphics.
Early work focused on laying out objects based on manually-defined design principles~\cite{ifurniture_design}, simple statistical relationships between objects extracted from a small set of examples~\cite{yu2011MakeItHome}, or with programmatically-specified constraints~\cite{rj_mcmc}.
Later research focused on data-driven methods~\cite{fisher2012,kermani2016learning,liang2017,qi2018human}, with a surge in activity as deep neural networks gained popularity~\cite{FastSynthCVPR,DeepSynthSIGGRAPH2018,wang2019planit,GRAINS,zhou2019scenegraphnet,zhang2019hybrid,Paschalidou2021NEURIPS,wang2020sceneformer,tang2023diffuscene}. 
These prior works develop closed-universe generative models (i.e. restricted to certain scene and object cateories), and all of them require (in some cases quite large) datasets of 3D scenes for training.
By contrast, LLMs offer the capability---in theory---to synthesize arbitrary types of scenes and to do so with no explicit training data.

\vspace{-0.5em}
\paragraph*{Scene Synthesis with LLMs}
While research on text-based scene generation predates the rise of LLMs~\cite{WordsEye,chang2017sceneseer3dscenedesign}, their development has led to a new generation of text-to-scene generative models which are both more flexible and more open-ended than earlier systems.
While LayoutGPT~\cite{feng2023layoutgpt} generated layouts by explicitly specifying the coordinates of objects, the rest of the early approaches have followed the declarative approach, i.e. using an LLM to produce a declarative program specifying the layout constraints~\cite{yang2023holodeck,SceneCraft,wen2023anyhome,aguinakang2024openuniverseindoorscenegeneration,çelen2024idesign,kodnongbua2024zeroshotsequentialneurosymbolicreasoning}.
Our work differs from these approaches, as we employ an imperative approach to scene synthesis, i.e. the LLM generates explicit instructions for constructing scenes by iteratively placing objects with relative positioning.
To the best of our knowledge, The Scene Language~\cite{zhang2024scenelanguagerepresentingscenes}, a concurrent work, is the only other approach which follows an imperative approach to LLM-based scene generation, though it tackles a different version of the problem focused on simpler scenes.
In addition, we present an apples-to-apples comparison between declarative and imperative approaches to scene layout generation, which we expect to be instructive for designing DSLs for LLM-based program synthesis on other tasks such as 3D shape modeling and editing.


\vspace{-0.5em}
\paragraph*{Correcting LLM Outputs}
As LLMs can fail to produce correct output in one shot, many prior works deploy corrective mechanisms to refine LLM-generated outputs, including some work on LLM-based scene generation~\cite{SceneCraft}. 
The most common approach is self-correction, where the output of an LLM is iteratively refined by the LLM itself~\cite{pan2023automaticallycorrectinglargelanguage}.
Such self-correction mechanisms, while appealing in theory, are costly to run (requiring multiple LLM calls), and on code generation tasks, they typically offer modest or no real performance gain~\cite{olausson2024repair}.
Instead, we propose an efficient error correction scheme based on iterative local search, finding programs that are close to the LLM's original output while minimizing errors such as object overlaps.

\vspace{-0.5em}
\paragraph*{Automated Evaluation Metrics for Text-based Visual Generation}
The surge of interest in text-based visual generation models also raises the question of how to evaluate such models.
While human perceptual studies are the `gold standard,' researchers have proposed other automated metrics based on multimodal large models (MLMs) for evaluating how well a generated visual asset respects an input text prompt.
For evaluating text-to-image models, approaches include CLIPScore~\cite{hessel2021clipscore}, VQAScore~\cite{lin2024evaluating}, and Davidsonian Scene Graphs (DSG)~\cite{JaeminCho2024}; for text-to-3D models, a method based on GPT-4V has been proposed~\cite{wu2023gpteval3d}.
For the types of complex 3D scenes we consider in this paper, we found that prior approaches were not well-aligned with human preferences.
Thus, we propose a simple new automated evaluation scheme that uses a single call to a multimodal LLM.
To the best of our knowledge, there are no prior automated evaluation methods designed specifically for text-to-3D-scene generation.

\section{Overview}

\begin{figure*}[t!]
    \centering
    \includegraphics[width=0.9\linewidth]{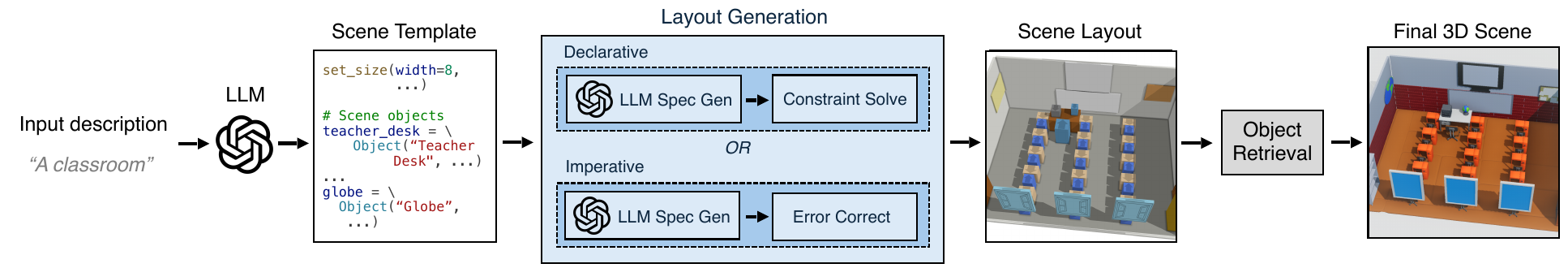}
    \caption{
    Our scene synthesis pipeline.
    An LLM first converts an input text description into a scene template (scene dimensions and list of objects).
    Then, a layout generation stage determines the positions and orientations of those objects using either the declarative or imperative paradigm.
    Finally, an optional object retrieval stage determines 3D meshes for each object in the scene.
    Regardless of the choice of layout generation method in blue, all other stages not shown in blue are kept fixed for fair comparison between layout methods.
    }
    \label{fig:overview}
\end{figure*}

Our goal is to investigate LLM-generated \emph{imperative} layouts for 3D scene synthesis and to compare them to \emph{declarative} layouts.
As much as possible, we would like this comparison to show the relative merits of these overall \emph{paradigms} for layout generation, rather than details of how a particular scene synthesis system was implemented.
Thus, we design a scene synthesis pipeline which factors out computational stages not relevant to layout generation and shares those stages in common between the different layout generation methods that we consider.

Figure~\ref{fig:overview} shows an overview of this pipeline.
In the first stage, an LLM takes a textual scene description as input and outputs a ``scene template'' consisting of the dimensions of the scene and a list of objects.
Each object is defined by a name, a set of dimensions, and one of three types of physical support: \texttt{STANDING}, \texttt{WALL-MOUNTED}, or \texttt{FLOATING}.
This scene template is then passed to a layout generation stage to determine the positions and orientations of its objects.
If the layout generation stage uses the declarative paradigm, it has two sub-stages: first using an LLM to generate the declarative layout specification, and then invoking a solver to produce a layout consistent with that specification.
If the layout generation stage uses the imperative paradigm (i.e. our method), it also has two sub-stages: LLM-based layout specification generation followed by our iterative error correction mechanism to improve the generated layout.
Finally (and optionally), the pipeline can invoke an object retrieval stage to retrieve a 3D mesh for each objects in the layout.
Object retrieval is not the focus of our work, but we include it in our pipeline for visualizing some qualitative results.
In our implementation, we use simple CLIP similarity~\cite{clip} to retrieve a mesh whose rendered image matches the object's name.
We retrieve meshes from the HypeHype Asset Library~\cite{HypeHypeAssets}, which currently contains about 6000 3D model assets.
Other 3D shape datasets could also be used here~\cite{chang2015shapenet,objaverse,objaverseXL}.

By comparing different layout methods while keeping the template synthesis and object retrieval stages the same, we ensure that scenes being compared have the same size and the same set of objects, making the comparison fair. To be consistent with prior work, we restrict object orientations to four cardinal directions.

In the next sections, we introduce a representative declarative language for scene layout specification, our imperative language, and compare the two (Section~\ref{sec:languages}).
We next describe our error correction scheme for imperative layout generation (Section~\ref{sec:errorcorrection}) and then present experimental results (Section~\ref{sec:results}).

\section{Scene description languages}
\label{sec:languages}
In this section, we introduce two domain-specific languages (DSLs) for scene layout generation: a declarative DSL and an imperative DSL. Both DSLs are embedded in Python, enabling the use of expressive constructs such as loops and conditionals while aligning with the strengths of LLMs in Python code generation. We describe each DSL and conclude with a discussion of why the imperative DSL addresses key limitations of the declarative approach.

\subsection{Declarative Language for Scene Layouts}
\label{sec:declarative}

In the declarative paradigm, scene layouts are specified by defining relational constraints between objects, such as spatial relationships, alignments, or adjacency. These constraints are then processed by a solver to compute a scene layout that satisfies as many constraints as possible. Declarative DSLs streamline this process by offering abstractions for compactly defining relationships.

We adopt the DSL and gradient-based solver introduced in~\cite{aguinakang2024openuniverseindoorscenegeneration}. The DSL includes commands for specifying object relations such as \texttt{adjacent}, \texttt{aligned}, and \texttt{facing}, as summarized in Table~\ref{tab:declarative_dsl}. For example, the command \texttt{adjacent(nightstand, bed, WEST, NORTH)} constrains the nightstand to be west of the bed while aligned along its north side.
A complete program for specifying a garage scene is shown in Figure~\ref{fig:code-compare}, illustrating how the declarative DSL can be used to specify a scene.

The gradient-based solver operates by converting the relational constraints into differentiable loss functions, where each constraint contributes to the total loss based on how well it is satisfied. The solver then optimizes object positions and orientations to minimize this loss.
For additional implementation details, please refer to~\cite{aguinakang2024openuniverseindoorscenegeneration}.

\begin{table}[t!]
\begin{tabular}{p{\dimexpr\columnwidth-2\tabcolsep\relax}} 
\toprule
Declarative DSL Relations \\
\midrule
\small \texttt{on(a: Obj, b: Obj)} \\
\small \texttt{next\_to\_wall(a: Obj, wall: int, distance: float)} \\
\small \texttt{mounted\_on\_wall(a: Obj, wall: int, h: float, b: Obj)} \\
\small \texttt{mounted\_on\_ceiling(a: Obj, b: Obj)} \\
\small \texttt{adjacent(a: Obj, b: Obj, dir: int, distance: float)} \\
\small \texttt{adjacent(a: Obj, b: Obj, dir\_1: int, dir\_2: int)} \\
\small \texttt{aligned(cuboids: list[Obj], axis: int)} \\
\small \texttt{facing(a: Obj, b: Obj)} \\
\small \texttt{surround(chairs: list[Obj], table: Obj)} \\
\bottomrule
\end{tabular}
\caption{Relations in a declarative DSL for describing scenes (based on~\cite{aguinakang2024openuniverseindoorscenegeneration}). 
}
\label{tab:declarative_dsl}
\end{table}
\subsection{Imperative Language for Scene Layouts}
\label{sec:imperative}

\begin{table}[t!]
\centering
\begin{tabular}{l|l}
\toprule
\small \textbf{Explicit Geometric Relationships} & \small \textbf{Use of Variables} \\ 
\hline
\small \texttt{chair.max.x = table.min.x - 0.1} & \small \texttt{d = 1.0} \\
\small \texttt{chair.center.y = table.center.y} & \small \texttt{for i, c in enum(cols):} \\
\small \texttt{chair.min.z = scene.min.z} & \small \texttt{\quad c.center.x = \textbackslash } \\
\small \texttt{chair.facing = table} & \small \texttt{  scene.center.x +  i * d} \\
\bottomrule
\end{tabular}
\caption{Key features of the imperative DSL. The left column demonstrates explicit geometric relationships for positioning objects relative to others. The right column shows the use of variables to define reusable patterns, enabling concise scene descriptions. Together, these features allow the imperative paradigm to describe scenes precisely and efficiently.}
\label{tab:imperative-dsl-features}
\end{table}

In the imperative paradigm, scene layouts are generated by explicitly specifying object positions and orientations in a step-by-step manner. Unlike the declarative paradigm, which solves jointly for all object positions and orientations, the imperative approach incrementally defines each object’s position and orientation relative to other objects or the scene itself. This direct specification enables precise and flexible control over the layout.

The simplest imperative strategy, as used in LayoutGPT~\cite{feng2023layoutgpt}, places all objects with respect to the scene’s global bounds. However, this approach results in less coherent layouts because it ignores relationships between objects. Instead, our imperative DSL allows objects to be positioned relative to one another. Each object is endowed with attributes such as \texttt{min}, \texttt{max}, \texttt{center}, and \texttt{facing}, which can be used to describe positions compactly and intuitively. Table~\ref{tab:imperative-dsl-features} (left) demonstrates how these attributes simplify the specification of a chair's position and orientation.

Additionally, our DSL supports the creation and usage of variables, which reduce redundant definitions of numeric values and provide flexibility in specifying relationships. For instance, shared variables can be used to define repeated patterns or symmetrical arrangements, making the code both concise and adaptable. Table~\ref{tab:imperative-dsl-features} (right) shows the code to position columns along the $x$ axis with equally spaced parameterized distances between them.

Figure~\ref{fig:code-compare} (left) shows an example of an imperative DSL program for describing a garage and its resulting scene layout, illustrating how these features work together to specify complex arrangements. These features of the imperative DSL—explicit geometric constraints and parameterized variables—enable precise and flexible scene layouts, addressing key limitations of naive imperative strategies. By allowing direct specification of geometric relationships, the imperative approach supports diverse and customizable scene layouts while reducing ambiguity. In the next subsection, we discuss the strengths and limitations of the two paradigms.

\subsection{Comparing Scene Description Languages}
\label{sec:comparing}

The imperative paradigm offers several notable advantages over the declarative approach. Below, we discuss these benefits in detail.

\vspace{-0.5em}
\paragraph{Scene Specification}
The imperative paradigm provides greater flexibility for defining open-ended layouts that go beyond the constraints of a fixed DSL grammar, as in the case of the declarative paradigm. For example, arranging objects in a row with alternating types is straightforward in the imperative style using loops and conditional logic. In contrast, declarative systems require explicit relations for every layout pattern. Expanding a declarative DSL to handle more complex arrangements---such as introducing a new relation for alternations---quickly becomes infeasible as it increases the language’s complexity. By allowing explicit and direct geometric constraints, the imperative approach avoids these limitations and supports diverse, customized scene specifications.


\vspace{-0.5em}
\paragraph{LLM Prompting}  
The declarative paradigm often involves complex DSLs that must be carefully explained to the LLM through hand-crafted in-context examples. For complex DSLs, this can become a significant impediment, as longer documentation and intricate command definitions increase the likelihood of LLM confusion. Even a single function with multiple parameters may require numerous examples for the LLM to understand its correct usage. For instance, the \texttt{adjacent(a, b, dir\_1, dir\_2)} command (cf. Sec.~\ref{sec:declarative}) is frequently misunderstood, with the LLM confusing the meanings of \texttt{dir\_1} and \texttt{dir\_2}, even after extensive prompting. In contrast, the imperative approach eliminates this complexity. Instead of requiring detailed explanations of DSL commands, the LLM only needs a few examples demonstrating how scenes are constructed using explicit geometric relations. This simplicity reduces the chance of errors and makes prompting significantly more straightforward and reliable.

\vspace{-0.5em}
\paragraph{Error Correction}  
Errors in declarative systems can stem from either syntactic issues, which are relatively easy to address with resampling, or semantic errors, such as incorrect relations or their parameterization. Semantic errors are particularly problematic because they often arise from the LLM’s misunderstanding of the DSL and only become apparent after constraint solving, making them hard to diagnose and fix. Self-correction mechanisms, which rely on iterative LLM prompting, frequently fail to resolve such errors, as the underlying misunderstanding of the DSL tends to persist.  
In contrast, the imperative approach eliminates the need for constraint solving, making the generated scene directly amenable to symbolic analysis. This allows for a simple, LLM-free correction mechanism that performs local optimization on scene parameters to fix errors effectively. We introduce this mechanism in the next section.

\section{Layout Error Correction}
\label{sec:errorcorrection}


Scene layouts generated by LLMs often contain errors. While resampling can sometimes fix errors in simple scenes, it is insufficient for intricate and complex layouts. Correcting these scenes directly offers a more reliable and efficient solution.

Our focus is on semantic errors, as syntactic issues can often be resolved by simply resampling the program. Semantic errors arise when either incorrect relationships are created or relationships are misparameterized. 
By adopting the imperative style, we significantly reduce incorrect relationship errors, as it lowers the ambiguity in how scenes are specified. 
However, errors in scene parameterization, such as miscalibrated distances or sizes, still persist. For example, in a scene with columns arranged in a row (e.g., using the code in Table~\ref{tab:imperative-dsl-features}, right), an incorrect spacing value (\texttt{d}) may cause overlaps (if \texttt{d < column.size.x}) or out-of-bounds placements (if \texttt{d * len(cols) > room.size.x/2}). Our goal is to correct these layout parameterization errors without additional LLM queries.

To address parameterization errors, we introduce an iterative correction mechanism that starts with the scene generated by the LLM and incrementally refines it by adjusting its parameters. The goal is to minimize a loss function that quantifies scene errors, such as overlaps between objects or objects placed out-of-bounds, while preserving the structure and intent of the LLM’s original output.

To minimize the loss while preserving the original structure of the LLM-generated scene, we adopt a coordinate descent-inspired approach. Instead of adjusting multiple parameters simultaneously, which risks altering the scene significantly, we iteratively update one parameter at a time. Starting from the initial scene configuration, we evaluate variations of each individual parameter and identify the single adjustment that results in the greatest loss reduction. This ensures that each step of the correction process introduces minimal deviation from the original scene. Once the adjustment is made, the updated configuration becomes the new starting point, and the process repeats. The process stops when the loss improvement between successive steps falls below a predefined threshold $\epsilon$. Further details are provided in the supplementary material.

This approach is particularly advantageous for imperative scenes due to the use of shared parameters across objects, such as a single spacing parameter for a row of columns or chairs in a theater. 
Editing such parameters allow cohesive adjustments across multiple objects while maintaining the structural rules of the scene. 
By working in this lower-dimensional parameter space, the mechanism ensures aesthetic consistency while also avoiding the computational overhead of per-object adjustments.
This property, unique to imperative scene descriptions, makes error correction efficient, coherent, and well-suited for scenes with many interrelated objects---an advantage that is difficult to replicate in declarative systems reliant on independently specified constraints.

By iteratively refining parameters, our method efficiently corrects errors while preserving the intent of the LLM-generated scene. On average, only a few adjustments---$7.13$ per scene, on average---are sufficient to resolve errors, demonstrating the robustness and scalability of this approach for complex layouts.
See the supplemental material for videos illustrating the error correction process.

\begin{table}[t!]
    \renewcommand{\arraystretch}{0.8}
    \centering
    \small
    \begin{tabular}{rcc}
        \toprule
        \textbf{Scene Type} & \textbf{Ours vs. \declarativebaseline} & \textbf{Ours vs. Holodeck}
        \\
        \midrule
        Overall & 82.86\% & 94.29\%
        \\
        \midrule
        Small & 71.43\% & 100\%
        \\
        Medium & 82.86\% & 91.43\%
        \\
        Large & 90.48\% & 95.24\%
        \\
        \midrule
        Indoor & 81.25\% & 95.83\%
        \\
        Outdoor & 86.36\% & 90.91\%
        \\
        \midrule
        Realistic & 84\% & 92\%
        \\
        Fantastical & 80\% & 100\%
        \\
        \midrule
        Chaotic & 74.19\% & 96.77\%
        \\
        Structured & 89.74\% & 92.31\%
        \\
        \bottomrule
    \end{tabular}
    \caption{Preference rates for scenes generated using our imperative approach vs. two declarative approaches in a forced-choice perceptual study.}
    \label{tab:perceptualstudy}
\end{table}

\begin{table}[t!]
    \centering
    \small
    \begin{tabular}{cccc}
        \toprule
        \textbf{LLMCompare} & \textbf{LLMCompare (no +/-)} & \textbf{VQAScore} & \textbf{DSG} \\
        \midrule
        77.14\% & 70\% & 58.57\% & 50.71\%
        \\
        \bottomrule
    \end{tabular}
    \caption{How frequently different automated evaluation metrics agreed with the majority human judgment from our perceptual study.
    Our new method (LLMCompare) is simple and outperforms prior approaches designed for evaluating text-to-image generative models.}
    \label{tab:humanalign}
\end{table}

\section{Results and Evaluation}
\label{sec:results}

In this section, we evaluate different layout generation approaches on their ability to synthesize open-universe 3D scenes.
We compare our imperative approach to two declarative approaches using forced-choice perceptual studies.
We also compare using a new automated evaluation method, which we show is better aligned with the perceptual study results than previous methods for automatic evaluation of text-based visual generation systems.

\vspace{-0.5em}
\paragraph*{Implementation Details}
Unless otherwise specified, we use Anthropic's \texttt{claude-3-5-sonnet-20241022} for language generation components of our proposed system.
We use OpenAI's \texttt{gpt-4o} as the multimodal LLM backbone for  automated evaluation methods.

\vspace{-0.5em}
\paragraph*{Benchmark}
To evaluate layout generation methods, we created a benchmark of 70 scene prompts spanning diverse environments and complexity levels. Each prompt is labeled by four attributes: \emph{size} (small, medium, large), \emph{location} (indoor, outdoor), \emph{realism} (realistic, fantastical), and \emph{structure} (chaotic, structured).
The benchmark includes 14 small, 35 medium, and 21 large scenes; 48 indoor and 22 outdoor; 50 realistic and 20 fantastical; and 31 chaotic and 39 structured.
See the supplemental for the complete list of scene prompts.

\begin{table*}[t]
\centering
\small
\begin{tabular}{cccc|ccc|ccc|ccc}
\toprule
\multirow{2}{*}{\centering\textbf{Model}}& 
\multicolumn{3}{c}{\texttt{\small claude-3-5-sonnet-20241022}} &
\multicolumn{3}{c}{\texttt{\small gpt-4o-2024-11-20}} &
\multicolumn{3}{c}{\texttt{\small o1-2024-12-17}} &
\multicolumn{3}{c}{\texttt{\small gemini-exp-1206}} \\
\cmidrule(lr){2-13}
 & \textsc{All} & \textsc{Bound} & \textsc{Ovl} 
 & \textsc{All} & \textsc{Bound} & \textsc{Ovl} 
 & \textsc{All} & \textsc{Bound} & \textsc{Ovl} 
 & \textsc{All} & \textsc{Bound} & \textsc{Ovl} \\
\hline
\textbf{\small Ours w/o correction}  & 17.57 & 1.56 & 10.76 & 17.80 & 1.14 & 11.19 & 17.29 & 2.36 & 10.70 & 14.67 & 1.19 & 10.41 \\
\textbf{\small Ours} & \textbf{2.10}  & \textbf{0.54} & \textbf{0.56}  & \textbf{3.34}  & \textbf{0.61} & \textbf{0.64}  & \textbf{3.24}  & \textbf{0.63} & \textbf{1.17}  & \textbf{3.14} & \textbf{0.61} & \textbf{1.14} \\
\bottomrule
\end{tabular}
\caption{Comparison of coordination errors, out-of-bounds placements, and overlaps for different LLM configurations across three setups: explicit coordinate prediction, pre-correction, and post-correction.}
\label{tab:llm_results_restructured}
\end{table*}

\vspace{-0.5em}
\paragraph*{Comparison Conditions}
In the subsequent experiments, we compare the following LLM-based scene layout generation methods:
\begin{itemize}
    \item \textbf{Ours}: generating imperative layout specifications and then improving them using our iterative error correction scheme.
    \item \textbf{\declarativebaseline}: the declarative layout generation approach described in Section~\ref{sec:declarative}.
    \item \textbf{Holodeck}: the ``Constraint-based Layout Design Module''
    of the Holodeck system~\cite{yang2023holodeck}.
\end{itemize}




\subsection{Perceptual Study}
\label{sec:perceptual_study}

To compare how well different layout generation methods can produce layouts satisfying the scene prompts in our benchmark, we conducted two-alternative, forced-choice perceptual studies pitting our method against each of the declarative methods.
We recruited 10 participants from a population of university students. The participants were divided into two groups, one for each study.
Each participant was shown a series of 70 comparisons (one per scene prompt in our benchmark), where each comparison contains a scene prompt, images of two layouts in randomized order, and a question asking them to choose which scene they thought was better (taking into account overall scene plausibility and appropriateness for the prompt).
For each comparison, we take the majority vote across all participants as the final answer.
Since we seek to evaluate only the quality of object layouts, to eliminate any impact that 3D model choice might have on participant response, objects in images were rendered as colored boxes over which participants could hover their mouse cursor to reveal the object's name.

Table~\ref{tab:perceptualstudy} shows the results of this experiment, and Figure~\ref{fig:qual_compare} shows some qualitative comparisons between generated layouts.
Overall, participants preferred layouts generated using our imperative method to those generated by either of the declarative versions.
As Holodeck is so strongly dis-preferred overall, there are not obvious trends in how scene types correlate with preference.
In the comparison against \declarativebaseline, there is a bigger preference gap for larger and less chaotic scenes.
These results suggest that while the imperative approach is effective overall, it is especially well-suited for large, dense scenes with considerable structure in their layouts.

\subsection{Automated Evaluation}
As perceptual studies are costly to run, we also investigated using automated evaluation metrics to approximate the results of a perceptual study.
We experimented with two metrics designed for automated evaluation of text-to-image generative models:
\begin{itemize}
    \item \textbf{VQAScore}~\cite{lin2024evaluating}: Scores how well a generated image matches a text prompt using the probability a visual question answering model assigns to the output token `yes' when asked whether the image depicts that text prompt. 
    \item \textbf{Davidsonian Scene Graphs (DSG}~\cite{JaeminCho2024}: Computes a score by generating a dependency graph of simpler yes/no questions to ask about the image and then aggregating the percentage of those questions for which a VQA system returns 'yes.'
\end{itemize}
We can use these methods to compare two scene layouts by running them on rendered images of both and returning whichever has the higher score.
Unfortunately, we found that neither of these methods performed much better than chance (50\%) at agreeing with the majority-vote judgments from our perceptual study.
DSG, in particular, struggled to generate yes/no questions which could differentiate the two scenes, leading to ties for most judgments.

These results motivated us to develop a simple new method for automated evaluation of scene layouts.
Specifically, we prompt a multimodal LLM with images of two scene layouts (along with a general task prompt) and ask it to list the pros and cons of each layout with respect to the scene prompt.
At the end of its output, the LLM returns which scene is better.
As shown in Table~\ref{tab:humanalign}, this simple method is much more aligned with human judgments.
Table~\ref{tab:humanalign} also includes results for an ablated version of our method which does not ask the LLM to first generate a pros \& cons list, illustrating that this additional step does improve the method's agreement with people.

Running our automated evaluation metric on our benchmark results in our imperative scenes being chosen 77.14\% of the time over \declarativebaseline~scenes (vs. 82.86\% in the perceptual study) and 90\% of the time over Holodeck scenes (vs. 94.29\% from the perceptual study).
While there is some discrepancy from `gold standard' human judgments, the trends are still clear.

\subsection{Error Correction}

Table~\ref{tab:llm_results_restructured} reports how many imperative layout errors our error correction scheme fixes across our scene prompt benchmark using different LLM backbones.
Different LLMs may produce layout programs that are more or less suitable for error correction, based on how they parameterize the layout.
Claude 3.5 Sonnet's layouts are most amenable to correction, resulting in the fewest overall errors after correction is applied.
Interestingly, OpenAI's inference-time compute model o1 is not noticeably better than GPT-4o in this case.


\subsection{Timing}

The scene template generation stage of our pipeline takes 9.45s on average.
Generating a \declarativebaseline~layout program takes 10.01s, whereas genearting an imperative layout program takes 19.22s.
However, the \declarativebaseline~layout optimizer takes 21.3s, whereas our imperative error correction procedures takes 9.26s.
Overall, it takes 40.76s to synthesize a declarative scene vs. 37.93s for an imperative scene, i.e. the average runtimes of the two paradigms are comparable.

\section{Conclusion and Future Work}
\label{sec:conclusion}

We introduced a new method for open-universe scene layout generation that adopts an imperative paradigm, in contrast to the declarative approaches commonly used in prior work. Our imperative method simplifies scene specification by facilitating direct placement of objects with respect to existing objects. Additionally, we proposed an iterative, LLM-free error correction mechanism, which refines generated scenes by adjusting scene parameters to improve validity while remaining close to the original layout.
We evaluated our approach through a forced-choice perceptual study, showing that participants preferred scenes generated by our method over two declarative baselines 82\% and 94\% of the time, respectively. Finally, we also introduced a novel automated evaluation metric for judging scene layouts, demonstrating that (1) scenes generated by our method achieve higher scores compared to alternatives, and (2) this metric aligns more closely with human preferences than existing automated evaluation metrics.

\paragraph{Limitations}  
Despite its advantages, our method has certain limitations. First, object orientation in our pipeline is restricted to four cardinal directions. 
Extending this to support arbitrary orientations is necessary for handling a wider variety of scenes. 
Further, the error correction mechanism, although more efficient than self-correction methods in declarative systems, is not fully inexpensive and is limited to parametric adjustments. It cannot address errors stemming from incorrect object retrieval or misinterpretation of input text prompts. Addressing such errors may require a hybrid approach that integrates symbolic error correction, like ours, with LLM-supported self-correction mechanisms, and warrants further investigation.

\paragraph{Future Work}  
While scene generation has received significant attention in research, much of it has focused on static scenes. A natural next step is to extend scene synthesis to dynamic scenes, where objects interact or evolve over time. As the generation tasks grow in complexity, simplifying the coding process for LLMs, as done with our imperative approach, will become increasingly important.

One of the key contributions of this work is providing an apples-to-apples comparison of LLM coding capability with two contrasting DSL paradigms. This comparison highlights a novel consideration when designing DSLs: optimizing them for LLM usage. Traditionally, DSLs are designed with only human usage or system efficiency in mind, but as more real-world applications adopt LLM-based solutions, designing LLM-friendly DSLs will become essential. We hope this work inspires future research to develop guidelines for designing DSLs that balance usability, flexibility, and LLM compatibility.

\bibliographystyle{ACM-Reference-Format}
\bibliography{main}
\clearpage









\appendix
\section{Error Correction Mechanism}

We provide additional details about the error correction mechanism introduced in the main paper. Specifically, we describe: (1) how the loss function is calculated, (2) how parameters are modified during the optimization process, and (3) how ties are broken during coordinate descent.

\paragraph{Loss Function}  
To quantify the quality of a scene layout, we define a \emph{loss function}, $\text{loss}(L)$, which penalizes various layout errors. This function is composed of the following terms:
\begin{enumerate}
\item \textbf{Out-of-Bounds Loss}: For each object, this loss is the linear distance by which its bounding cuboid protrudes outside the scene boundary. If the object is fully within bounds, this term is zero.
\item \textbf{Overlap Loss}: For each pair of objects, this loss is the average linear size of the intersection of their bounding cuboids. To ensure usability, doors and windows are assigned expanded collision boxes to account for opening space and prevent obstruction.
\item \textbf{Standing Loss}: For each \texttt{STANDING} object, this term measures the distance from the bottom face of the object’s cuboid to the nearest horizontal surface supporting it.
\item \textbf{Mounted Loss}: For each \texttt{MOUNTED} object, this term is the distance between the object’s mountable face (usually the back or side) and the closest vertical surface.
\end{enumerate}
These components ensure that the loss function captures both structural integrity and functional correctness in the layout.

\paragraph{Parameter Modification during Optimization}  
The optimization process starts with the initial LLM-generated program, $P_{\text{LLM}}$, and iteratively refines it by constructing a sequence of programs, $P_0 = P_{\text{LLM}}, P_1, P_2, \ldots$. At each step, a set of candidate programs, known as the \emph{neighborhood} $N(P_i)$, is generated by applying predefined edits to the parameters of $P_i$.

The neighborhood $N(P_i)$ is formed by varying each floating-point constant or orientation expression in $P_i$ using a predefined set of modifications. Examples of such modifications include:
\begin{itemize}
\item Multiplying a floating-point constant by 2 or 0.5.
\item Adding or subtracting a small predefined offset to a parameter.
\item Flipping an orientation value (e.g., rotating by 90° or reversing a facing direction).
\end{itemize}

From this set of candidates, the program with the smallest loss is selected for the next step. This process ensures that only minimal changes are made at each step, preserving the structure of the original scene. The optimization stops when the loss reduction between successive steps falls below a predefined threshold, $\varepsilon$.

\paragraph{Tie-Breaking during Coordinate Descent}  
If multiple candidates in the neighborhood $N(P_i)$ have the same minimal loss, we use a tie-breaking mechanism to ensure consistency. Ties are resolved based on the \emph{mass transport distance}, which measures how much the updated layout deviates from the original.

The mass transport distance, $d_{\text{mass}}(L_1, L_2)$, between two layouts $L_1$ and $L_2$ is defined as:
\begin{equation}
d_{\text{mass}}(L_1, L_2) = \sum_{o \in O(L_1)} \text{Vol}(o) \cdot \|\text{center}(o) - \text{center}(f(o))\|,
\end{equation}
where $O(L_1)$ and $O(L_2)$ are the sets of objects in $L_1$ and $L_2$, respectively, and $f$ is a bijection matching objects between the two layouts. Larger objects are penalized more for movement, encouraging the preservation of the positions of major scene elements while allowing smaller objects to be adjusted.

If two candidates are still tied after considering the mass transport distance, a simple deterministic rule (e.g., index-based ordering) is used to select one. This tie-breaking mechanism ensures that changes to the scene are minimal and consistent, avoiding unnecessary disruptions to its structure.

These additional details highlight how the error correction mechanism systematically reduces layout errors while maintaining the structural intent of the original scene. By balancing structural fidelity, computational efficiency, and robustness, our approach ensures that the corrected layouts are both functional and coherent.

\section{Holodeck Modifications}
To focus on evaluating the quality of object layouts and eliminate any influence object selection might have on participant responses in the perceptual study, Holodeck’s object selection module was modified to use only the same set of objects and sizes present in the corresponding scenes from Ours/DeclBase. To avoid prompting the LLM for the object selection plan json, which resulted in hallucinations of objects and object sizes outside of the given constraints, a 'mock' json following the LLM output format was manually created and inserted into the system pipeline for the layout module to use. Because the original system makes the LLM simultaneously select the objects and specify secondary object relations (specifying which of the objects are on top of other objects), removing this module resulted in the Holodeck scenes with missing secondary objects for the perceptual study, such as computers lying on the floor next to desks rather than on top.


\begin{figure*}[ht!]
    \centering
    \includegraphics[width=0.9\linewidth]{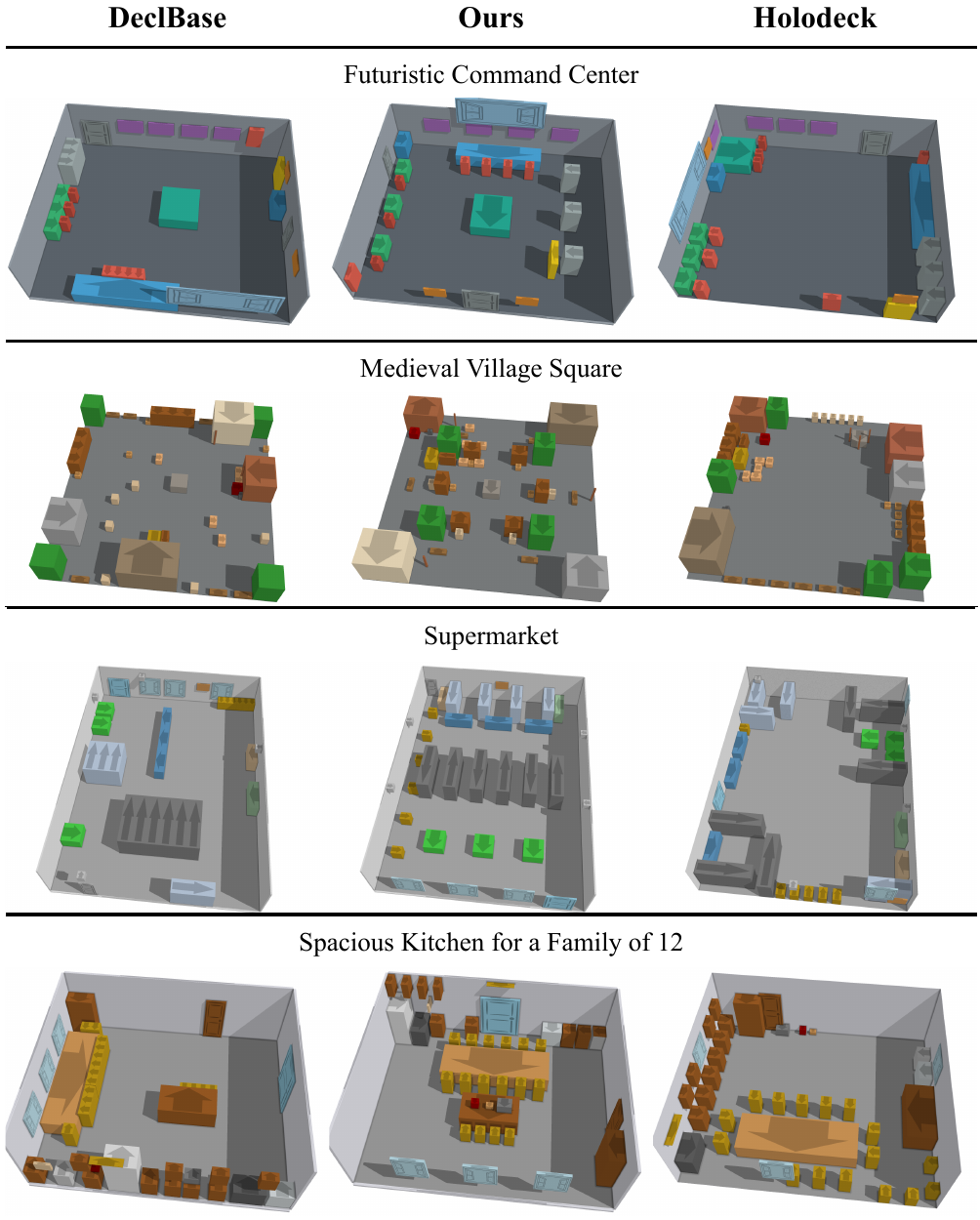}
    \caption{
    Qualitative comparisons between our method, DeclBase, and Holodeck. Our method and Holodeck uses \texttt{gpt-4o}, while DeclBase uses \texttt{claude-3-5-sonnet-20241022}. See the supplemental for a comparison between our method and DeclBase only using \texttt{claude-3-5-sonnet-20241022}.
    }
    \label{fig:qual_compare}
\end{figure*}

\begin{figure*}[ht!]
    \centering
    \includegraphics[width=\linewidth]{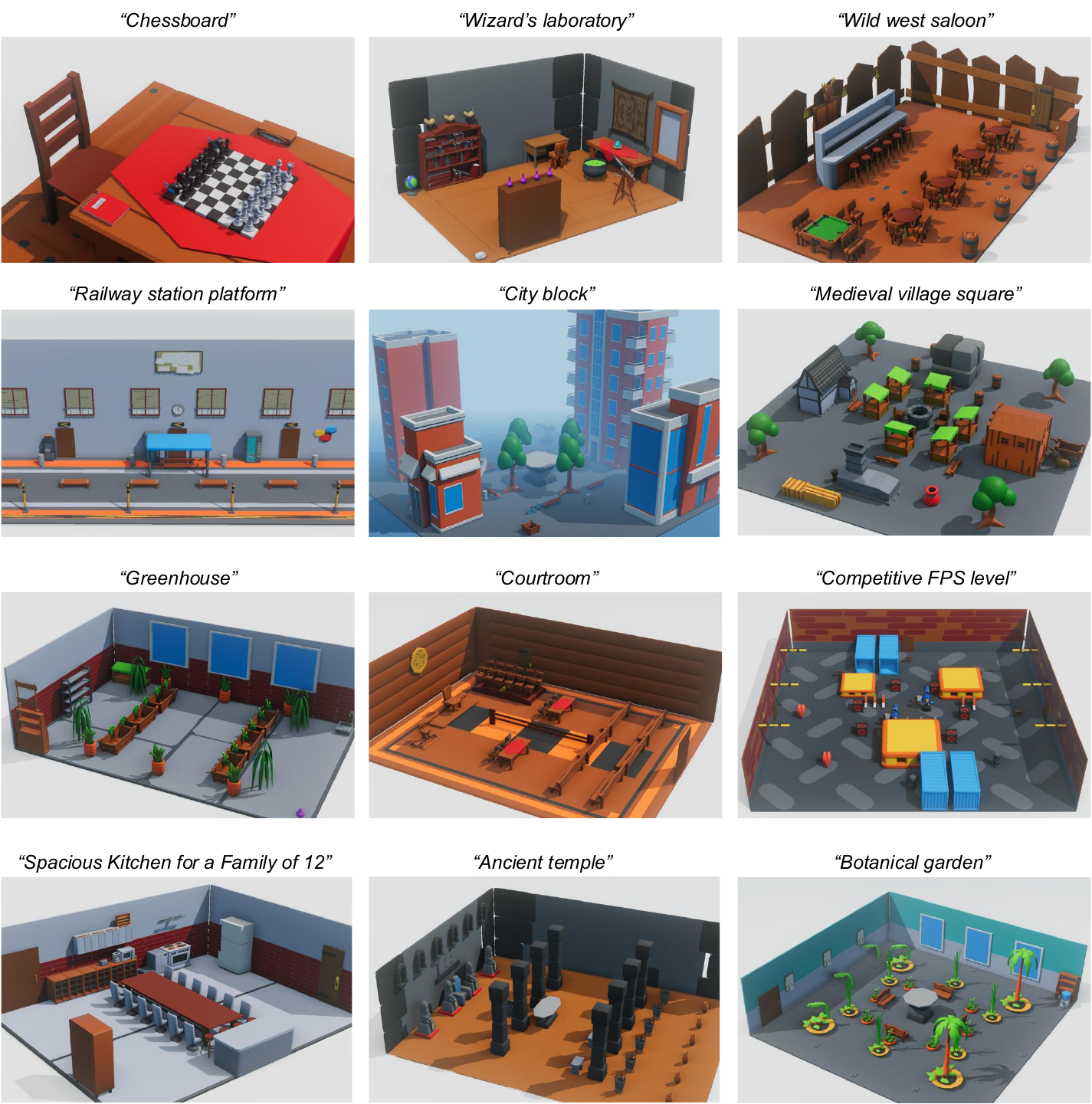}
    \caption{
    More scenes synthesized using our imperative layout generation method with error correction.
    }
    \label{fig:qualitative_hh}
\end{figure*}

\end{document}